\documentclass[12pt,a4paper,oneside,final,tilepage,onecolumn,thmsal]{article}
\usepackage[dvips]{graphics}
\usepackage{epsfig}
\usepackage{amsmath}

\newcommand{\QED}{\hspace*{\fill}\rule{2.5mm}{2.5mm}}

\usepackage{graphics}
\begin{document}
\def\beq{\begin{equation}}
\def\eeq{\end{equation}}
\def\bea{\begin{eqnarray}}
\def\eea{\end{eqnarray}}
\def\ve{\vert}
\def\vel{\left|}
\def\ver{\right|}
\def\nnb{\nonumber}
\def\ga{\left(}
\def\dr{\right)}
\def\aga{\left\{}
\def\adr{\right\}}
\def\rar{\rightarrow}
\def\nnb{\nonumber}
\def\la{\langle}
\def\ra{\rangle}
\def\ba{\begin{array}}
\def\ea{\end{array}}
\def\ds{\displaystyle}
\title{{\small {\bf The asymptotic iteration method for the angular spheroidal
 eigenvalues with arbitrary complex size parameter $c$  }}}
\author{\vspace{1cm}\\
{\small T. Barakat\thanks {electronic address:
zayd95@hotmail.com}, K. Abodayeh, B. Abdallah, and O. M. Al-Dossary}\\
{\small Physics Depntartment, King Saud University}\\
{\small Riyadh, Saudi Arabia }}
\date{}
\begin{titlepage}
\maketitle
\thispagestyle{empty}
\begin{abstract}
\baselineskip .8 cm The asymptotic iteration method is applied, to
calculate the angular spheroidal eigenvalues
$\lambda^{m}_{\ell}(c)$ with arbitrary complex size parameter
$c$. It is shown that, the obtained numerical results of
$\lambda^{m}_{\ell}(c)$ are all in excellent agreement with the
available published data over the full range of parameter values
$\ell$, $m$, and $c$. Some representative values of
$\lambda^{m}_{\ell}(c)$ for large real $c$ are also given.
\end{abstract}
\vspace{1cm} PACS number(s): 02.70.-c.
\end{titlepage}
\section{{\small Introduction}}
\baselineskip .8cm \hspace{0.6cm} The solution of the spheroidal
wave equation is a very old subject, but it still an important
theme in the existing literature. The importance of this equation
arises in many areas of physics. For instance, it plays an
important role in the study of light scattering in optics [1-3],
nuclear modeling [4], signal processing and communication theory
[5], electromagnetic modeling [6], and in finding the
electromagnetic induction (EMI) response of canonical objects at
magnetoquasistatic frequencies [7]. Applications utilizing complex
$c$ include for example, light scattering from spheroidal
particles, and spheroidal antennas enveloped in a plasma medium.

Attempts to find rapid, and accurate eigenvalues
$\lambda^{m}_{\ell}(c)$ of the angular spheroidal wave equation
for large size parameter $c^2$ (assumed real) have been ongoing.
Flammer summarizes the work (up to 1957 [8]), and documents the
asymptotic expansions for $\lambda^{m}_{\ell}(c)$ of the angular
spheroidal wave equation. Since that time, serious attempts for
this case were made by many authors. Slepian [9], and Streifer
[10] derived uniform asymptotic expansions for the spheroidal
functions and their eigenvalues, which were further developed by
des Cloiseaux, and Mehta [11], and Dunster [12]. Other asymptotic
results based on WKB methods have been obtained by Sink, and Eu
[13]. Recently, asymptotic expansions of $\lambda^{m}_{\ell}(c)$
 for large $l$, and $c$ have
been proposed by Guimar\"{a}es [14], de Moraes and Guimar\"{a}es
[15], and the work of Do-Nhat [16, 17] summarizes and provides
more details of Flammer's expansions for $\lambda^{m}_{\ell}(c)$.

Nevertheless, because of the complexity of the angular spheroidal
wave equation with arbitrary complex size parameter $c = c_{r} +
c_{i}i$, where $c_{r} =Re\{c\}$, and $c_{i} =Im\{c\}$, evaluation
of $\lambda^{m}_{\ell}(c)$ in this regime has been much less
studied.

Very recently, Barrowes et al. [18] compute the asymptotic
expansions of $\lambda^{m}_{\ell}(c)$ with arbitrary complex size
parameter $c$ in the asymptotic regime of large $\mid c\mid$ with
$l$, and $m$ fixed. On the other hand, few packages have been
developed for the computation of the angular spheroidal
eigenvalues $\lambda^{m}_{\ell}(c)$ with arbitrary complex size
parameter $c = c_{r} + c_{i}i$. Thompson [19], Li et al. [20,
 21], and Falloon et al. [22] are of the most recent ones.

Those attempts to obtain the eigenvalues $\lambda^{m}_{\ell}(c)$
with arbitrary complex size parameter $c = c_{r} + c_{i}i$ are
rely heavily on power series expansions, and complicated
recurrence relations. Accurate results in those works are
obtainable at the expense of extensive mathematical, and
numerical manipulations, thus obscuring the physical analysis of
the corresponding system.

The present work applies the asymptotic iteration method (AIM)
[23, 24], for the computation of the angular spheroidal
eigenvalues $\lambda^{m}_{\ell}(c)$ with arbitrary complex size
parameter $c = c_{r} + c_{i}i$. This method was applied by
Barakat et al. [25] to compute the angular spheroidal eigenvalues
$\lambda^{m}_{\ell}(c)$ with real $c^2$, and for the
eigenenergies of the anharmonic oscillator potential [26]. The
implementation of this method was straightforward, and the
results were sufficiently accurate for practical purposes. Most
importantly, the numerical computation of the angular spheroidal
eigenvalues using this method was quite simple, fast, and the
eigenvalues were satisfying a simple ordering relation.
Therefore, one can unambiguously select the correct starting
eigenvalue.

Furthermore, AIM was quite flexible in the sense that, it is
applicable to any parameter value involved like $\ell$, $m$, and
$c$. It also handles $\lambda^{m}_{\ell}(c)$ with large $\ell$,
and $c$ which poses many numerical instabilities to some of the
previously mentioned methods. Therefore, the main motivation of
the present work is to overcome the shortcomings of those
approaches, and to formulate an elegant algebraic approach to
yield a fairly simple analytic formula which will give rapidly
the eigenvalues with high accuracy.

In this spirit, this paper is organized as follows. In Sec. 2 the
asymptotic iteration method for the angular spheroidal wave
equation is outlined. The analytical expressions for asymptotic
iteration method are cast in such a way that allows the reader to
use them without proceeding into their derivation. In Sec. 3 we
present our numerical results compared with other works, and then
we conclude and remark therein.

\section{{\small Formalism of the asymptotic iteration method for the
angular spheroidal wave equation}}
 \hspace{0.6cm}
 The angular spheroidal wave equation, with which we shall be
 concerned, is
\begin{eqnarray}                    
\frac{d}{d\eta}\left[(1-\eta^2)\frac{d}{d\eta}S_{\ell,m}(c;\eta)
\right]+ \left[
(\lambda^{m}_{\ell}(c))^2-c^2\eta^2-\frac{m^2}{(1-\eta^2)}\right]
 S_{\ell,m}(c;\eta)= 0.
\end{eqnarray}

The parameter $c$, which is related to the ellipticity of the
spheroidal coordinate surfaces, is allowed to be a complex
variable in the present work. Consequently, there is no need to
make the usual distinction between the prolate, and oblate forms
of the spheroidal wave equation, and the prolate form equation (1)
is adopted for definiteness. The other parameters
$\lambda^{m}_{\ell}(c)$, and $m$ are separation constants.

The second arises as a wave number for the polar angle of
spheroidal coordinates and, as usual, is required to be a
nonnegative integer; $\ell\geq m$ is an integer enumerating the
eigenvalues, and functions.

The spheroidal wave functions $S_{\ell,m}(c;\eta)$ are defined to
be the solutions of equation (1) that are finite at the two end
points $\eta=\pm1$ of the range of the independent variable. These
finiteness can be satisfied only for certain eigenvalues
$\lambda^{m}_{\ell}(c)$, which depend on the values of $c$ once a
specific value of  $m$ has been chosen.

The simplest case is that of $c$ = 0, for which the function
$S_{\ell,m}(c;\eta)$ reduces to the associated Legendre function,
and $(\lambda^{m}_{\ell}(c))^2=\ell(\ell+1)$ is its eigenvalues.

Here, the integer $\ell$ labels successive eigenvalues for fixed
$m$. When $\ell = m$ we have the lowest eigenvalue, and the
corresponding eigenfunction has no nodes in the interval
$-1\leq\eta\leq 1$. When $\ell = m + 1$ we have the next
eigenvalue, and the eigenfunction has one node inside (-1, 1);
and so on. A similar situation holds for the general case $c\neq
0$.

In order to apply the AIM for the general case $c\neq 0$, we have
to investigate the behavior of the solution near the singular
points $\eta=\pm1$. Substituting a power series expansion of the
form
\begin{eqnarray}                         
S_{\ell,m}(c;\eta)=(1-\eta^2)^{\alpha}\sum_{k=0}^{\infty}a_{k}(1-\eta^2)^{k},
\end{eqnarray}
into equation (1), we find that the regular solution has $\alpha=
m/2$. Without loss of generality, we can take $m\geq 0$ since
$m\rightarrow -m $ is a symmetry of the equation. Therefore, we
get an equation that is more tractable to the method if we factor
out this behavior. Accordingly, we set
\begin{eqnarray}                             
S_{\ell,m}(c;\eta)=(1-\eta^2)^{m/2} y_{\ell,m}(c;\eta),
\end{eqnarray}
then the new function $y_{\ell,m} (c;\eta)$ will satisfy a
second-order homogenous linear differential equation of the form
\begin{eqnarray}                              
(1-\eta^2)\frac{d^2y_{\ell,m}
(c;\eta)}{d\eta^2}-2(m+1)\eta\frac{dy_{\ell,m} (c;\eta)}{d\eta}+
(\varepsilon-c^2\eta^2)y_{\ell,m} (c;\eta)=0,
\end{eqnarray}
where
\begin{eqnarray}                            
\varepsilon\equiv(\lambda^{m}_{\ell}(c))^2-m(m+1).
\end{eqnarray}

Both equations (1), and (4) are invariant under the replacement
$\eta\rightarrow -\eta$. Thus the functions $S_{\ell,m}(c;\eta)$,
and $y_{\ell,m} (c;\eta)$ must also be invariant, except possibly
for an overall scale factor.

The systematic procedure of the AIM begins by rewriting equation
(4) in the following form
\begin{eqnarray}                               
y_{\ell,m}^{''} (c;\eta)=\lambda_{0}(\eta)y_{\ell,m}^{'}
(c;\eta)+s_{0}(\eta)y_{\ell,m} (c;\eta),
\end{eqnarray}
where
\begin{eqnarray}                                 
\lambda_{0}(\eta)=\frac{2(m+1)\eta}{(1-\eta^2)},~ and
~~~~s_{0}(\eta)=-\frac{\varepsilon-c^2\eta^2}{(1-\eta^2)},
\end{eqnarray}
and following the technique of AIM [23, 25], that will lead to a
general solution of equation (6):
\begin{eqnarray}                                  
y_{\ell,m} (c;\eta)=exp\left(-\int^{\eta}\beta(\eta^{'})
d\eta^{'}\right)\left[C_{2}+C_{1}\int^{\eta}
exp\left(\int^{\eta^{'}}\{\lambda_{0}(\eta^{''})+2\beta(\eta^{''})\}d\eta^{''}\right)d\eta^{'}\right].
\end{eqnarray}
If for some $n>0$,
\begin{eqnarray}                                    
\beta(\eta)\equiv\frac{s_n(\eta)}{\lambda_n(\eta)}=\frac{s_{n-1}(\eta)}{\lambda_{n-1}(\eta)},
\end{eqnarray}
with
\begin{eqnarray}                                    
\lambda_{n}(\eta)=\lambda^{'}_{n-1}(\eta)+s_{n-1}(\eta)+\lambda_{0}(\eta)\lambda_{n-1}(\eta),
~and~~~
s_{n}(\eta)=s^{'}_{n-1}(\eta)+s_{0}(\eta)\lambda_{n-1}(\eta).
\end{eqnarray}

For sufficiently large $n$, we can now introduce the termination
condition of the method, which in turn, yields the angular
spheroidal eigenvalues $\lambda^{m}_{\ell}(c)$
\begin{eqnarray}
\delta_{n}(\eta)\equiv
s_{n}(\eta)\lambda_{n-1}(\eta)-s_{n-1}(\eta)\lambda_{n}(\eta)=0.
\end{eqnarray}

\section{{\small Numerical results for the angular spheroidal eigenvalues $\lambda^{m}_{\ell}(c)$}}

Within the framework of the AIM mentioned in the above section,
the angular spheroidal eigenvalues $\lambda^{m}_{\ell}(c)$ are
calculated by means of equation (11). To obtain the eigenvalues
$\lambda^{m}_{\ell}(c)$, the iterations should be terminated by
imposing a condition $\delta_{n}(\eta)$= 0 as an approximation to
equation (11). On the other hand, for each iteration, the
expression
$\delta_{n}(\eta)=s_{n}(\eta)\lambda_{n-1}(\eta)-s_{n-1}(\eta)\lambda_{n}(\eta)$
 depends on two variables: $\lambda^{m}_{\ell}(c)$, and $\eta$. The
calculated eigenvalues $\lambda^{m}_{\ell}(c)$ by means of this
condition should, however, be independent of the choice of
$\eta$. Nevertheless, the choice of $\eta$ is observed to be
critical only to the speed of the convergence to the eigenvalues,
as well as for the stability of the process. In this work it is
observed that, the best starting value for $\eta$ is the value at
which the effective potential of equation (1) takes its minimum
value. For this purpose, it is necessary to perform the variable
change $\eta \rightarrow tanh(x)$, mapping the finite interval
(-1,1) into the infinite one $(-\infty,\infty )$, and then
equation (1) can be rewritten as
\begin{eqnarray}
-\frac{d^2}{dx^2}S_{\ell,m}+V_{eff}(x)S_{\ell,m}=-m^2S_{\ell,m},
\end{eqnarray}
where the effective potential $V_{eff}(x)$ is
\begin{eqnarray}
V_{eff}(x)=-\left [(\lambda^{m}_{\ell}(c))^{2}-c^2\right]
sech^2(x)-c^{2}sech^4(x).
\end{eqnarray}
$V_{eff}(x)$ is an even function, its minimum value occurs when
$x=0$, which in turn implies that $\eta=0$. Therefore, at the end
of the iterations we put $\eta=0$.

To test the rate of convergence of AIM numerically, we calculate
the angular spheroidal eigenvalue $\lambda^{0}_{0}(10)$ shown in
figure 1. We simply tried a set of iteration numbers $n=5,
10,.....$, and the convergence of AIM seems to take place
smoothly when $n \geq 45$ iterations.

Proceeding in the same way, the results of the AIM for
$\lambda^{m}_{\ell}(c)$ with different values of $\ell$, $m$, and
$c$ are reported in tables 1, 2, 3, 4, and 5. The results are
shown in such a way that one can easily judge the accuracy of the
method. The angular spheroidal eigenvalues
$\lambda^{m}_{\ell}(c)$ were calculated by means of 45 iterations
only. Our calculated eigenvalues $\lambda^{m}_{\ell}(c)$ are all
in excellent agreement with the available published data [8, 18,
20, 22] over the full range of parameter values, $\ell$, $m$, and
$c$.

A second, more stringent, test of the method is shown in Table 3,
where we consider large real $c$, in this case, in order to
reproduce more accurate results, the angular spheroidal
eigenvalues $\lambda^{m}_{\ell}(c)$ were calculated by means of
100 iterations. Again the agreement is quite excellent with the
available published data.

It is worthwhile to emphasize that, the AIM is very easy to
implement for calculating the angular spheroidal eigenvalues
$\lambda^{m}_{\ell}(c)$, without having to worry about the
ranges, and forms of $c$. Moreover, for purely real c, and purely
imaginary c, the obtained eigenvalues are satisfying a very simple
ordering relation. But, for complex arbitray c values, the
eigenvalues are no longer real, since the spheroidal equation is
not self-adjoint. In this case the ordering is determined by the
absolute magnitude of $(\lambda^{m}_{\ell}(c))^2$. Hence, one can
unambiguously select the correct starting eigenvalue. This
represents a significant advantage over the tridiagonal matrix
method [22] in which, the eigenvalues are not ordered, and hence
to choose the correct matrix eigenvalue, one must use an
iterative process to move towards the starting value.

 As a concluding remark, we would like to point out that, the accuracy
of the results could be increased if the number of iterations are
increased.
\\

This work is supported by King Saud University, College of Science
- Research center projects No. (PHYS/2005/26).

\clearpage
\begin{table}
\begin{center}
\caption{Comparison of selected values of eigenvalues
$(\lambda^{m}_{\ell}(c))^2$ computed by Flammer [8], Le-Wei Li et
al. [20], and by means of the present work.}
 \vspace{1cm}
\begin{tabular}{ccccc}
\hline \hline \multicolumn{1}{c}{}&\multicolumn{4}{c}{
$(\lambda^{m}_{\ell}(c))^2$}\\\cline{3-5}
$c^2$ & (m,$\ell$) & Flammer [8]&Le-Wei Li et al. [20] &Present work \\
\hline
 -1.00 & (4,11)  & 131.560& 131.560 &131.560 \\
 \hline
  0.10& (2,2)    & 6.01427& 6.01427 & 6.01427\\
  \hline
  1.00  &(1,1)   &2.19555& 2.19555 &2.19555  \\
        &(2,2)   &6.14095& 6.14095 &6.14095\\
        &(2,5)   &30.4361& 30.4362 & 30.4362\\
 \hline
 4.00   &(1,1)   &2.73411& 2.73411 &2.73411\\
        &(2,2)   &6.54250& 6.54250 &6.54250 \\
\hline
  16.00& (1,1)   &4.39959& 4.39959 &4.39959\\
       & (2,5)   &36.9963& 36.9963 &36.9963\\

\hline\hline
\end{tabular}
\end{center}
\end{table}

\begin{table}
\begin{center}
\caption{Comparison of selected values of eigenvalues
$(\lambda^{m}_{\ell}(c))^2$ computed by Falloon et al. [22], and
by means of the present work.} \vspace{1cm}
\begin{tabular}{cccc}
\hline \hline \multicolumn{1}{c}{}&\multicolumn{3}{c}{
$(\lambda^{m}_{\ell}(c))^2$}\\\cline{2-4}
c & (m,$\ell$) &Falloon et al. [22]&Present work \\
\hline
10  & (0,0)   &9.228304     &9.228304 \\
    & (0,1)   &28.13346     &28.13346 \\
    & (0,2)   &45.86895     &45.86895  \\
    & (1,1)   &10.28777     &10.28777 \\
    & (1,2)   &29.33892     &29.33892  \\
    & (1,3)   &47.30152     &47.30152  \\
    & (2,2) & 13.46308     &13.46308 \\
    & (2,3) & 32.93818    &32.93818   \\
    & (2,4) & 51.52485    &51.52485   \\
\hline
10$i$ & (0,0) & -81.02794    &-81.02794 \\
      & (0,1) &-81.02794     & -81.02794\\
      & (0,2) &-45.48968     &-45.48968  \\
      & (1,1) &-62.11935     &-62.11935 \\
      & (1,2) &-62.11915     &-62.11915  \\
      & (1,3) &-29.18576     &-29.18576  \\
      & (2,2) &-43.29025     & -43.29025 \\
      & (2,3) &-43.28716     &-43.28716   \\
      & (2,4) &-13.50811     &-13.50811   \\

\hline \hline\hline
\end{tabular}
\end{center}
\end{table}
\newpage

\begin{table}
\begin{center}
\caption{Comparison of selected values of eigenvalues
$(\lambda^{m}_{\ell}(c))^2$ computed by J. W. Liu [27], Falloon
et al. [22], and by means of the present work.} \vspace{1cm}
\begin{tabular}{ccccc}
\hline \hline \multicolumn{1}{c}{}&\multicolumn{4}{c}{
$(\lambda^{m}_{\ell}(c))^2$}\\\cline{3-5}
c & (m,$\ell$) & J. W. Liu [26]&Falloon et al. [22]&Present work \\
\hline

50  & (0,0)   &49.24615   &-           &49.24615\\
    & (0,1)   &148.2306   &-           &148.2306\\
    & (0,3)   &343.1109   &-           &343.1109 \\
    & (0,4)   &438.9725   &-           &438.9725  \\
    & (1,1)   &50.25646   &-           &50.25646\\
    & (1,2)   &149.2622   &-           &149.2622 \\
    & (1,4)   &344.1894   &-           &344.1894 \\
    & (1,5)   &440.0769   &-           &440.0769 \\
\hline
100 & (0,0)   &99.24810   &99.24810    &99.24810\\
    & (0,1)   &298.2405   &298.2405    &298.2405\\
    & (0,2)   &-          &496.2212    &496.2212  \\
    & (0,3)   &693.1825   &-           &693.1825  \\
    & (0,4)   &889.1162-  &-           &889.1162 \\
    & (1,1)   &100.2532   &100.2532    &100.2532 \\
    & (1,2)   &299.2558   &299.2558    &299.2558\\
    & (1,3)   &-          &497.2472    &497.2472\\
    & (1,4)   &694.2195   &-           &694.2195 \\
    & (1,5)   &890.1645   &-           &890.1645 \\
\hline\hline
\end{tabular}
\end{center}
\end{table}
\newpage

\begin{table}
\begin{center}
\caption{Comparison of selected values of eigenvalues
$(\lambda^{m}_{\ell}(c))^2$ computed by Le-Wei Li et al. [20],
and by means of the present work. }
 \vspace{1cm}
\begin{tabular}{cccc}
\hline \hline \multicolumn{1}{c}{}&\multicolumn{3}{c}{
$(\lambda^{m}_{\ell}(c))^2$}\\\cline{3-4}
c & (m,$\ell$) & Le-Wei Li et al. [20]&Present work  \\
\hline
1.824770+2.601670$i$  &(0,0) &1.701836+4.219998$i$  &1.701836+4.219998$i$  \\
2.094267+5.807965$i$  &(0,2) & 1.993901+8.576325$i$ &1.993901+8.576325$i$  \\
5.217093+3.081362$i$  &(0,2) & 23.91023+18.74194$i$ &23.91033+18.74184$i$ \\
3.563644+2.887165$i$  &(0,1) & 10.13705+11.12216$i$ &10.13705+11.12218$i$  \\
1.998555+4.097453$i$  &(1,1) & 2.919098+6.134851$i$ &2.919095+6.134851$i$ \\
3.862833+4.492300$i$  &(1,1) & 12.19691+16.24534$i$ &12.19691+16.24534$i$ \\
2.136987+5.449457$i$  &(2,0) & 6.098946+7.684379$i$ & 6.098961+7.684333$i$ \\
\hline\hline
\end{tabular}
\end{center}
\end{table}

\newpage
\begin{table}
\begin{center}
\caption{Comparison of selected values of eigenvalues
$(\lambda^{m}_{\ell}(c))^2$ computed by B. E. Barrowes et al.
[18], and by means of the present work.} \vspace{1cm} \small
\begin{tabular}{cccc}
\hline \hline \multicolumn{1}{c}{}&\multicolumn{3}{c}{
$(\lambda^{m}_{\ell}(c))^2$}\\\cline{3-4}
c & (m,$\ell$) & B. E. Barrowes et al. [18]&Present work  \\
\hline
1.824770749208805+ 2.601670692890318$i$&(0,0)& 1.705180+4.220186$i$&1.705180+4.220186$i$  \\
3.563644553545243+ 2.887165344336900$i$&(0,1)&10.14084+11.12159$i$&10.14084+11.12159$i$  \\
5.217093042404772+ 3.081362886557631$i$&(0,2)&23.91583+18.74332$i$&23.91583+18.74332$i$ \\
4.067274712533398+ 6.264358978587767$i$&(0,3)&11.78093+22.54139$i$&11.78093+22.54139$i$  \\
2.244329796261236+ 8.973752190228394$i$&(0,4)&2.156125+12.81092$i$&2.156124+12.81092$i$ \\
7.606334073445308+ 6.906465157219409$i$&(0,5)&48.80665+54.01199$i$&48.80665+54.01197$i$ \\
6.316233767329015+ 9.949229739353585$i$&(0,6)&29.84604+55.68999$i$&29.84604+55.68998$i$ \\
\hline
1.998555442181652+ 4.097453662365392$i$&(1,0)& 2.915319+6.133951$i$& 2.915319+6.133951$i$  \\
3.862833529248772+ 4.492300074953849$i$&(1,1)& 12.20110+16.24408$i$&12.20110+16.24408$i$  \\
2.184204069300826+ 7.326156812534641$i$&(1,2)&3.102506+10.53921$i$ &3.102506+10.53921$i$ \\
7.270040170458184+ 5.010809182556227$i$&(1,3)& 47.41099+39.42618$i$&47.41098+39.42619$i$  \\
6.119087892218941+ 8.234638882858787$i$&(1,4)&29.88062+46.18128$i$ &29.88062+46.18128$i$ \\
4.510843794687041+11.068777156965684$i$&(1,5)&14.19982+38.58584$i$&14.19982+38.58584$i$ \\
\hline
2.136987377094029+ 5.449457313914277$i$&(2,0)&6.102540+7.684764$i$ &6.102540+7.684763$i$  \\
4.105156484215650+ 5.922750658440496$i$&(2,1)&16.13687+20.40623$i$ & 16.13687+20.40623$i$  \\
5.907125751703487+ 6.283464345814330$i$&(2,2)& 32.08759+34.48753$i$& 32.08758+34.48754$i$ \\
7.634658702877481+ 6.576768822063829$i$&(2,3)&53.56137+49.63852$i$ & 53.56136+49.63853$i$  \\
6.337223309080594+ 9.739915760209533$i$&(2,4)&34.18267+53.57326$i$ & 34.18267+53.57326$i$ \\
\hline
2.254441944326160+ 6.731940814252908$i$&(3,0)&11.27374+9.046369$i$ & 11.27374+9.046369$i$  \\
4.312789375877335+ 7.267942971679416$i$&(3,1)& 21.98994+24.06425$i$ & 21.98994+24.06425$i$  \\
6.178462421804212+ 7.686866389128842$i$&(3,2)&38.86809+40.69989$i$ &38.86808+40.69989$i$ \\
7.954733051349589+ 8.033437876318695$i$&(3,3)&61.43614+58.62467$i$ &61.43613+58.62466$i$  \\
\hline
2.357663561227191+ 7.971913317957412$i$&(4,0)&18.43306+10.28552$i$ &18.43306+10.28552$i$  \\
4.496463440238013+ 8.560827522001995$i$&(4,1)&29.78282+27.39486$i$  &29.78282+27.39486$i$  \\
6.420513778117898+ 9.029346203504799$i$&(4,2)& 47.52324+46.35817$i$&47.52324+46.35817$i$ \\
\hline
2.450444507315414+ 9.182664291501323$i$&(5,0)&27.58314+11.43648$i$ &27.58314+11.43648$i$  \\
4.662342487692823+ 9.817718575239088$i$&(5,1)& 39.52918+30.48922$i$&39.52918+30.48922$i$  \\
\hline
2.535162563188484+10.371846133322299$i$&(6,0)&38.72574+12.51971$i$&38.72574+12.51971$i$\\
 \hline\hline
\end{tabular}
\end{center}
\end{table}
\pagebreak
\begin{figure}
\centering
\includegraphics[width=0.8\textwidth, angle=0]{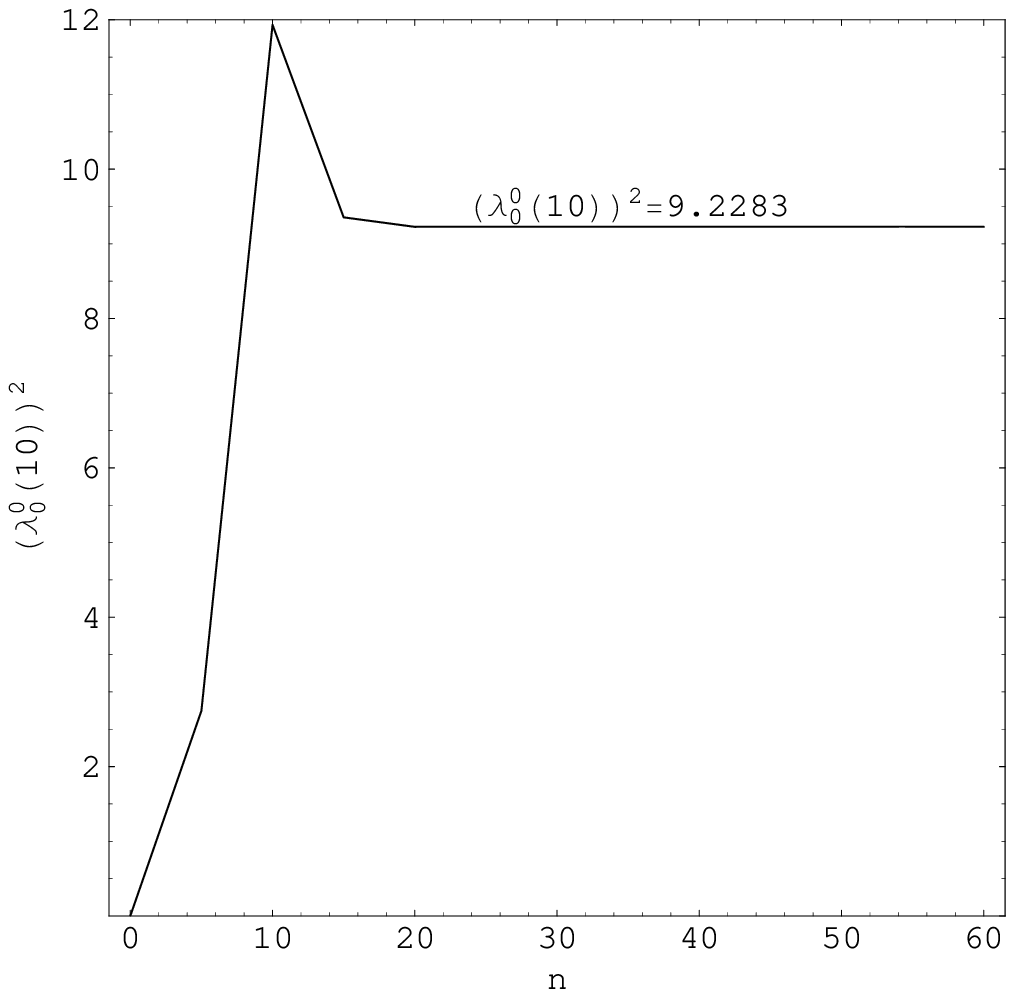}
\caption{The rate of convergence of AIM for $\lambda^{0}_{0}(10)$
as a function of the number of iterations $n$.} \label{fig 1}
\end{figure}

\clearpage
\begin{thebibliography}{99}
\bibitem{R1}S. Asano and M. Sato, Appl. Opt. {\bf 19}, 962
(1980).
\bibitem{R2} N. V. Voshchinnikov and V. G. Farafonov,
Astrophys. Space Sci. {\bf 204}, 19 (1993).
\bibitem{R3}M. I. Mishchenko, J. W. Hovenier, and L. D. Travis,
Light scattering by nonspherical particles: theory, measurements,
and applications. Academic Press (2002).
\bibitem{R4}B. D. B. Figueiredo, J. Phys. A {\bf 35}, 2877 (2002).
\bibitem{R5}B. Larsson, T. Levitina, and E. J. Brandas,
Int. J. Quan. Chem. {\bf 85}, 392 (2001).
\bibitem{R6}M. F. R. Cooray, I. R. Ciric, and B. P. Sinha,  Can. J.
Phys. {\bf 68}, 376 (1990).
\bibitem{R7}C. O. Ao, H. Braunisch, and K. O'Neill, IEEE Trans. on
Geoscience and Remote Sensing {\bf 40}, 887 (2002).
\bibitem{R8}C. Flammer, Spheroidal wave functions. Stanford:
Stanford University Press (1957).
\bibitem{R9}D. Slepian, J. Mah. and Phys. {\bf 44}, 99 (1965).
\bibitem{R10}W. Streifer, J. Math. and Phys. {\bf 47}, 407 (1968).
\bibitem{R11}J. des Cloiseaux and M. L. Mehta, J. Math. Phys. {\bf 13},
1745 (1972).
\bibitem{R12}T. M. Dunster, SIAM J. Math. Anal. {\bf 17},
1495 (1986).
\bibitem{R13}M. L. Sink and B. C. Eu, J. Chem. Phys. {\bf 78},
4886 (1983).
\bibitem{R14}L. G. Guimar\"{a}es, J. Phys. A {\bf 28}, L233
(1995).
\bibitem{R15}P. C. G. de Moraes and L. G. Guimar\"{a}es, Quant. Spect. Radia. Trans. {\bf
74}, 757 (2002).
\bibitem{R16}T. Do-Nhat, Can. J. Phys. {\bf 77}, 635 (1999).
\bibitem{R17}T. Do-Nhat, Can. J. Phys. {\bf 79}, 813 (2001).
\bibitem{R18}B. E. Barrowes, K. O'Neill, T. M. Grzegorczyk, and J.
A. Kong, Studies in Appl. Math. {\bf 113}, 271 (2004).
\bibitem{R19}W. J. Thompson, Comput. Sci. Eng. {\bf 1}, 84 (1999).
\bibitem{R20}L. W. Li, M. S. Leong, T. S. Yeo, P. S. Kooi, and K. Y. Tan, Phys. Rev. E {\bf 58},
6792 (1998).
\bibitem{R21}L. W. Li, X. K. Kang, and M. S. Leong, Spheroidal and Coulomb Spheroidal Functions (New York:
Wiley) (2002).
\bibitem{R22}P. E. Falloon, P. C. Abbott, and J. B. Wang, J. Phys.
A {\bf 36}, 5477 (2003);
www.physics.uwa.edu.au/~falloon/spheroidal/.
\bibitem{R23}H. Ciftci, R. L. Hall, and N. Saad, J. Phys.
A {\bf 36}, 11807 (2003); J. Phys. A {\bf 38}, 1147 (2005).
\bibitem{R24}F. M. Fern$\acute{a}$ndez, J. Phys.
A {\bf 37}, 6173 (2004).
\bibitem{R25}T. Barakat, K. Abodayeh, and A. Mukheimer, J. Phys.
A {\bf 38}, 1299 (2005).
\bibitem{R26}T. Barakat, Phys. Lett A {\bf 344}, 411 (2005).
\bibitem{R27}J. W. Liu,  J. Math. Phys. {\bf 33}, 4026 (1992).

\end{thebibliography}
\end{document}